\documentclass[aps,pra,reprint,twocolumn,floatfix]{revtex4-1}
\newcommand{\ket}[1]{\left\vert #1 \right\rangle}
\newcommand{\bra}[1]{\left\langle #1 \right\vert}

\usepackage{amssymb}
\usepackage{bm}
\usepackage{multirow,amsbsy,amsmath}
\usepackage{graphicx}
\usepackage{verbatim}
\usepackage{hyperref}
\usepackage{pifont}
\usepackage{color}
\usepackage{subfigure}
\usepackage{dcolumn}
\usepackage{hyperref}
\hypersetup{colorlinks=true, citecolor=blue, urlcolor=blue, linkcolor=blue}
\makeatother
\makeatletter

\begin{document}

\title{Nonclassical photon pair source based on the noiseless photon echo}
\author{Duo-Lun Chen}
\author{Zong-Quan Zhou}
\email{zq\_zhou@ustc.edu.cn}
\author{Chuan-Feng Li}
\email{cfli@ustc.edu.cn}
\author{Guang-Can Guo}

\affiliation{CAS Key Laboratory of Quantum Information, University of Science and Technology of China, Hefei 230026, China}
\affiliation{CAS Center for Excellence in Quantum Information and Quantum Physics, University of Science and Technology of China, Hefei 230026, China}
\affiliation{Hefei National Laboratory, University of Science and Technology of China, Hefei 230088, China}

\date{\today}

\begin{abstract}
	The Duan-Lukin-Cirac-Zoller (DLCZ) scheme is a potential method to establish remote entanglements and realize large-scale quantum networks.
	Here, we propose a DLCZ-like scheme based on the noiseless photon echo in rare-earth-ion doped crystals.
	Correlated photon pairs with a controllable delay can be created by the direct optical rephasing.
	Theoretical analysis indicates that the protocol is efficient in the low optical depth regime.
	This protocol could be feasibly implemented to establish long-lived quantum correlations between a photon and a spin-wave excitation in rare-earth-ion doped crystals.
\end{abstract}

\maketitle

\section{introduction}
The establishment of a strong quantum correlation between photons and spin waves via the Duan-Lukin-Cirac-Zoller (DLCZ) approach \cite{DLCZ} plays a vital role in the construction of transportable quantum memories \cite{zhong2015optically,morton2015spin,ma2021one} and memory based quantum networks \cite{sangouard2011quantum,kimble2008quantum}.
In the DLCZ scheme, a coherent laser pulse induces an emission of Stokes photons.
The detection of a Stokes photon establishes a quantum spin-photon correlation, which can be read out later via the detection of an anti-Stokes photon.
Remote entanglement can be established by further encoding the Stokes-anti-Stokes photon pairs on certain degrees of freedom \cite{chou2007functional,yuan2008experimental,choi2010entanglement,yu2020entanglement}.
Such DLCZ-like photon pair sources have been demonstrated in various physical systems, including atomic ensembles \cite{kuzmich2003generation,albrecht2015controlled,chou2007functional,yuan2008experimental,choi2010entanglement,yu2020entanglement,Li:21}, phonons in diamond \cite{lee2012macroscopic,PhysRevLett.111.243601}, mechanical resonators \cite{riedinger2016non}, and rare-earth ion doped crystals (REICs) \cite{beavan2012demonstration,ledingham2012experimental,laplane2017multimode,ferguson2016generation,kutluer2017solid,kutluer2019time}.

REIC is an attractive platform because of its wide bandwidth \cite{sun2012optical} and long coherence lifetimes for both optical \cite{equall1994ultraslow,thiel2011rare} and hyperfine \cite{ranvcic2018coherence,zhong2015optically} transitions.
However, it is challenging to create an efficient correlated photon pair via direct implementation of the DLCZ protocol in REICs, because of their weak dipole moments and the dephasing caused by large optical inhomogeneous broadenings \cite{ottaviani2009creating,Goldschmidt:13}.
To overcome this problem, two types of DLCZ-like protocols have been proposed for REICs and both are inspired by rephasing techniques developed in protocols for absorptive quantum memories. 
One is the rephased amplified spontaneous emission (RASE) \cite{RASE,beavan2012demonstration}, inspired by the two-pulse echo \cite{hahn1950spin} and the four-level echo (4LE) \cite{beavan2011photon}.
The other is the atomic frequency comb based DLCZ (AFC-DLCZ) protocol \cite{AFC-DLCZ}, inspired by the AFC quantum memory protocol \cite{de2008solid,afzelius2009}.

RASE uses a $\pi$ pulse to invert the atomic populations and produce the amplified spontaneous emission (ASE), corresponding to the Stokes emission in DLCZ. 
For the two-level RASE scheme, one must work in the low optical depth regime to avoid multiple photon events in ASE detection, which leads to a trade-off between the efficiency and the single photon purity \cite{Stevenson2014}.
The four-level RASE (4L-RASE) scheme avoids this trade-off by generating the photon pair via different transitions \cite{beavan2012demonstration,Stevenson2014},
but the distortion of the control pulses and the indirect spontaneous emissions via intermediate crystal field levels lead to noise \cite{beavan2012demonstration,ferguson2016generation}.
Experiments based on RASE have only enabled the generation of entanglements in the continuous variable regime thus far \cite{ferguson2016generation}.

Experiments based on AFC-DLCZ have demonstrated the generation of temporal multimode correlated photon pairs \cite{kutluer2017solid,laplane2017multimode} and time-bin entanglements \cite{kutluer2019time} in the discrete variable regime, where the rephasing is achieved by tailoring the natural absorption into a comb-like profile \cite{AFC-DLCZ}.
The consumption of the natural absorption could hinder the implementation of the AFC-DLCZ scheme in the optically thin regime, such as the europium ion dopant in yttrium orthosilicate (Eu$^{3+}$:Y$_{2}$SiO$_{5}$), which is an important material for quantum applications due to its 6-hour hyperfine coherence lifetime and 1-hour optical storage lifetime under a strong magnetic field \cite{zhong2015optically,ma2021one}.
However, the oscillator strength of Eu$^{3+}$ ions is typically weak \cite{konz2003temperature,PhysRevLett.128.180501} and going to strong magnetic fields could further reduce the available optical depth \cite{ma2021one}. 
The further reduction of the natural absorption caused by the spectral tailoring could be unbearable.

Recently, we proposed and demonstrated a noiseless photon echo (NLPE) \cite{ma2021elimination} protocol for absorptive quantum memories. The spontaneous emission noise and the coherent noise can be eliminated by double rephasing in the four-dimensional atomic Hilbert space. 
The noise of NLPE is much less than that of standard photon echoes, while the efficiency is better than that of AFC because of the preservation of the natural absorption.
Here we propose an NLPE-based DLCZ approach (NLPE-DLCZ). 
This protocol could generate photon pairs with higher nonclassical correlations in the optically thin medium, which can be suitable for implementations in Eu$^{3+}$:Y$_{2}$SiO$_{5}$ crystals.

The rest of this paper is organized as follows.
In Sec. \ref{sec1} we briefly outline the correlated protocols.
In Sec. \ref{sec2} the protocol of NLPE-DLCZ is introduced in detail.
In Secs. \ref{sec3} and \ref{sec4} we analyze the Stokes and anti-Stokes emissions and calculate the readout efficiency.
In Sec. \ref{sec5} we estimate the noise and the cross correlation.
In Sec. \ref{sec6} the temporal multimode capacity and the lifetime are evaluated. 
In Sec. \ref{sec7} the applications of NLPE-DLCZ are discussed.
The conclusion is drawn in Sec. \ref{sec8}.
Finally, in the Appendixes, some detailed calculations
omitted in the main text are presented.

\section{review on correlated protocols}
\label{sec1}

\begin{figure}[h!]
	\centering
	\includegraphics[width=0.48\textwidth]{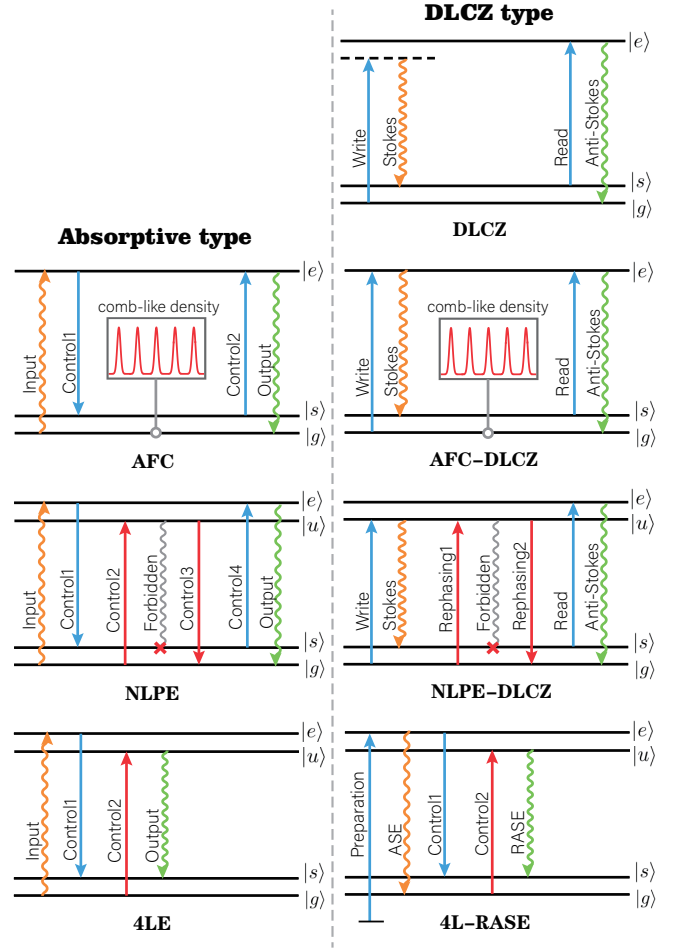}
	\caption{The pulse sequence for various absorptive memories and DLCZ-type protocols. All protocols initialize atoms in $\ket{g}$, except 4L-RASE prepares them in $\ket{e}$.
	In DLCZ, a off-resonant write pulse induces Raman transitions and produces a spin wave via a Stokes emission. The spin wave is read out by a resonant $\pi$ pulse via an anti-Stokes emission.  
	In AFC, the atomic spectral density is tailored to be comb-like, so as to rephase the coherence and produce the output echo. 
	The control fields are used to create and read out the spin wave. 
	AFC-DLCZ changes to create the spin wave via the Stokes emission induced by a resonant weak write pulse.
	In NLPE, the rephasing is achieved by two pairs of $\pi$ control pulses. The induced coherent noise of the $\pi$ pulses and the spontaneous emission from $\ket{u}$ can be filtered in spectral, which endows NLPE with low noise properties.
	NLPE can be converted to NLPE-DLCZ when the spin wave is created by the Stokes emission instead of a control pulse. 
	In 4LE, a pair of frequency-staggered $\pi$ pulses is applied to rephase the input field and generate an amplified echo, while in 4L-RASE these control pulses are used to rephase the ASE from $\ket{e}$ and produce the RASE.
	} 
	\label{pic0}
\end{figure}

Prior to introducing NLPE-DLCZ, it would be advantageous to present an overview of the absorptive optical memories and DLCZ-like protocols in REICs. The pulse sequence diagrams for these protocols are illustrated in Fig. \ref{pic0}.
Our starting point is the DLCZ-type correlated photon pair source \cite{DLCZ}, depicted in Fig. \ref{pic0}. 
The atoms are initialized in the ground state $\ket{g}$. 
An off-resonant write pulse induces Raman transitions into state $\ket{s}$,
thereby creating a spin wave via a Stokes emission.
Later, a $\pi$ pulse is employed to read this spin wave out via an anti-Stokes emission.
The key aspect of this process is the correlation between each pair of photons, which enables the realization of long-distance quantum communication through entanglement operations on these photon pairs \cite{DLCZ, chou2007functional,yuan2008experimental,choi2010entanglement,yu2020entanglement}.

When applying DLCZ to REICs, the off-resonant write pulse is typically substituted with a resonant one to provide sufficient excitation.
But optical inhomogeneities in REICs entail additional rephasing techniques to address the dephasing during resonant excitation  \cite{ottaviani2009creating,Goldschmidt:13}.
AFC is a well-known rephasing technique for REICs \cite{afzelius2009, de2008solid}. 
A typical AFC quantum storage is depicted in Fig. \ref{pic0}. 
The spectral density of the atoms initialized in $\ket{g}$ is tailored into a comb-like structure. 
Then the input optical field absorbed by the atoms forms coherence on the $\ket{g}-\ket{e}$ transition.
Due to the optical inhomogeneous broadening, the phase of the coherence for each atom evolves asynchronously. 
At a certain moment, the phases in adjacent teeth of the atomic frequency comb will differ exactly by $2\pi$ and the coherence is in phase due to the periodicity of phase, inducing an output echo.
To control the emitting time of the echo on demand, a pair of $\pi$ control pulses are introduced, which realizes the mutual transfer between the optical excitation and the spin-wave excitation \cite{jobez2015coherent, liu2020demand}.

AFC-DLCZ is a modified DLCZ protocol for REICs that can be obtained by transforming the AFC protocol above \cite{AFC-DLCZ}.
In the AFC storage, the spin wave is created by transferring the optical excitation via the $\pi$ control pulse, while in AFC-DLCZ it is generated by a weak resonant write pulse via the Stokes emission. 
In this way, the output field of AFC-DLCZ becomes an anti-Stokes emission and is correlated with the Stokes emission \cite{AFC-DLCZ, kutluer2017solid,laplane2017multimode}.

Photon echo is another important rephasing technique \cite{hahn1950spin, kurnit1964observation}. 
It achieves rephasing by inverting the atoms to upper energy state with optical pulses and forcing the phase of coherence evolve in reverse.
However, it is usually hard to apply photon-echo based protocol for quantum storage due to the noise issues \cite{ruggiero2009two}. 
The NLPE protocol is proposed to solve this problem \cite{ma2021elimination}. 
As depicted in Fig. \ref{pic0}, the rephasing is achieved by two pair of $\pi$ pulses.
The vital design of NLPE is to introduce an upper energy level $\ket{u}$ different from the level $\ket{e}$ that emits the output photons.
It ensures that all the control $\pi$ pulses are spectrally resolvable with the output field, so that the induced coherent noise can be filtered in spectral.
Moreover, by forbidding the $\ket{s}-\ket{u}$ transition, the imperfection of the $\pi$ pulses can only result in populations on $\ket{u}$ rather than $\ket{e}$, and thus the spontaneous emission noise from $\ket{u}$ can also be eliminated by spectral filtering.
These endows NLPE with low-noise properties \cite{ma2021elimination}.

The successful transformation from AFC to AFC-DLCZ inspires us to apply a similar replacement to NLPE, which leads to the invention of the NLPE-DLCZ protocol (see Fig. \ref{pic0}).
A thorough theoretical analysis in the following will demonstrate that this new protocol obtained by analogy is indeed effective.

Both AFC-DLCZ and NLPE-DLCZ establish the coherence with a weak write pulse. 
The coherence will vanish if the write pulse is replaced by a $\pi$ pulse.
Differently, the RASE type protocols \cite{RASE, beavan2012demonstration} employ a $\pi$ pulse to initialize the atoms into an upper energy state $\ket{e}$, and the coherence is created by the ASE from $\ket{e}$.
Photon echo techniques, such as using one control $\pi$ pulse (Hahn echo \cite{hahn1950spin}) or two frequency-staggered $\pi$ pulse (4LE \cite{beavan2011photon}), can be employed to rephase the coherence and produce a RASE \cite{RASE,beavan2012demonstration}. 
Thus, RASE can be regard as the `echo' of the corresponding ASE to some extent, which explains the correlation between them.
These correlated ASE-RASE pairs can also be used to apply DLCZ-based approaches in REICs \cite{beavan2012demonstration, ledingham2012experimental, ferguson2016generation}.

\section{protocol}
\label{sec2}

\begin{figure}[hbt!]
	\centering
	\subfigure[]{
		\includegraphics[width=0.4\textwidth]{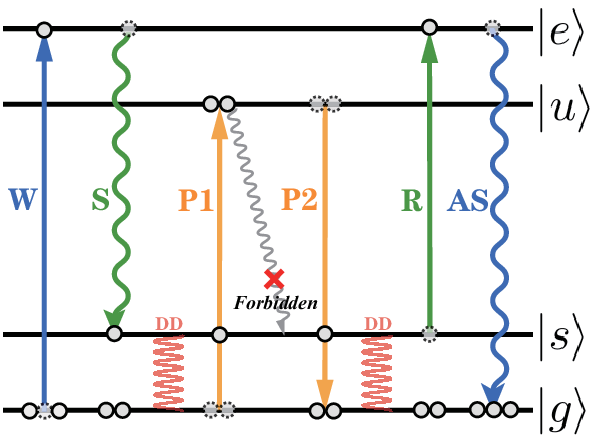}
		\label{1a}
	}
	\subfigure[]{
		\includegraphics[width=0.48\textwidth]{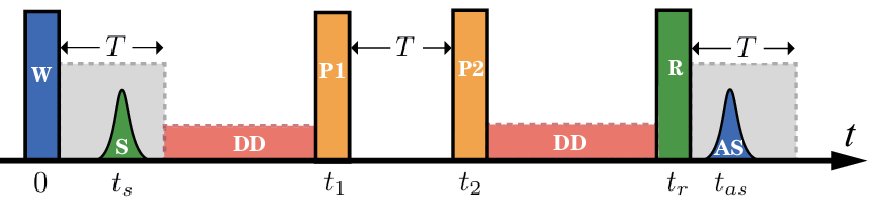}
		\label{1b}
	}
	\subfigure[]{
		\includegraphics[width=0.48\textwidth]{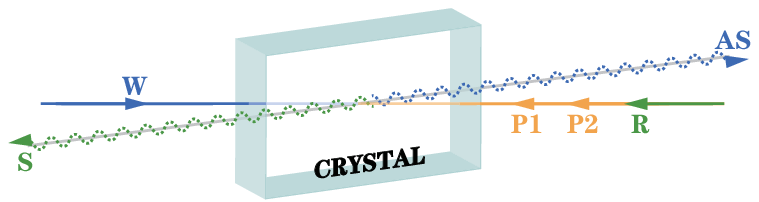}
		\label{1c}		
	}
	\caption{Pulse sequence of the NLPE-DLCZ scheme in the (a) frequency
		and (b) temporal domains to create a correlated photon pair. Circles in (a) represent the atomic populations. Initially, all atoms are prepared in $\ket{g}$. At time $t=0$, a resonant write pulse (W) induces a small excitation in $\ket{e}$, which may spontaneously emit Stokes photons during a time window of length $T$. The detection of a Stokes photon (S) at time $t_{S}$ heralds a spin-wave excitation in $\ket{s}$. After Stokes detection, a pair of rephasing $\pi$ pulses (P1 \& P2) incident at times $t_{1}$ and $t_{2}$, with their frequencies resonant with the $\ket{g}-\ket{u}$ transition. A resonant $\pi$ read pulse (R) at time $t_{r}$ excites the spin wave to $\ket{e}$ again. The excitation is efficiently mapped into an anti-Stokes photon at time $t_{AS}=t_{r}-t_{S}+T$ when the coherence is rephased. 
		To reduce noise, the $\ket{s}-\ket{u}$ transition is forbidden.
		Additionally, DD can be applied during $t\in[T,t_{1}]$ or $t\in[t_{2},t_{r}]$ to extend the lifetime of the quantum correlation. 
		(c) A recommended spatial configuration. 
		The rephasing beams (P1 \& P2) are in the same spatial modes as the read beam (R), while counterpropagating with the write beam (W). 
		The Stokes and anti-Stokes modes (S \& AS) counterpropagate with each other and have a small angle with the other beams. 
		The phase matching condition for the anti-Stokes emission $\bm{k}_{AS}+\bm{k}_{S}=\bm{k}_{w}-\bm{k}_{1}+\bm{k}_{2}+\bm{k}_{r}$ is satisfied.
	} 
	\label{pic1}
\end{figure}

In this section we formally introduce the NLPE-DLCZ protocol.
The pulse sequence of NLPE-DLCZ in the frequency and temporal domains are depicted in Figs. \ref{1a} and \ref{1b}, respectively.
The protocol is applied for an inhomogeneously broadened four-level atomic system.
Initially, all four-level atoms are initialized to the ground state $\ket{g}$ by an optical preparation procedure \cite{ma2021elimination}. 
Then at time $t=0$, a weak write pulse (W) with the wave vector $\bm{k}_{w}$ and an optical area $\theta_{0}\ll 1$ excites a small fraction of atoms to the excited state $\ket{e}$ and produces coherence on the $\ket{g}-\ket{e}$ transition.
The phase of the coherence evolves until the excitation spontaneously decays into another lower energy state $\ket{s}$ via a Stokes emission at time $t=t_{S}$. 
A time window of length $T$ is opened to detect the Stokes photons with the wave vector $\bm{k}_{S}$ and a successful detection heralds a spin-wave excitation on the $\ket{g}-\ket{s}$ transition.
Due to the dephasing caused by the optical inhomogeneous broadening, a single read pulse is inefficient to read the spin wave out and an extra rephasing procedure is required \cite{ottaviani2009creating,Goldschmidt:13}.

Inspired by the NLPE protocol \cite{ma2021elimination}, we achieve the rephasing by introducing a pair of $\pi$ pulses (the rephasing pair, P1 \& P2). Both pulses are resonant with the $\ket{g}-\ket{u}$ transition,  
where $\ket{u}$ is another excited state.
The rephasing pulses are sent at times $t_{1}$ and $t_{2}=t_{1}+T$ with wave vectors $\bm{k}_{1}$ and $\bm{k}_{2}$, respectively.
Then by sending a resonant $\pi$ pulse (the read pulse, R) with the wave vector $\bm{k}_{r}$ at time $t_{r}$, the phase of the $j$th atom at time $t_{AS}=t_{r}-t_{S}+T$ is
\begin{align}
\phi_{j}=&\phi_{j0}-\omega_{ge}^{j} t_{S}-\omega_{gs}^{j}(t_{1}-t_{S})
+\omega_{su}^{j}T\notag\\
&-\omega_{gs}^{j}(t_{r}-t_{2})-\omega_{ge}^{j}(t_{AS}-t_{r})  \notag\\
=&\phi_{j0}-\omega_{ue}^{j}T-\omega_{gs}^{j}(t_{r}-t_{S}),
\label{phase}
\end{align}
where $\phi_{j0}$ is the spatial phase and $\omega_{pq}^{j}$ $(p,q=g,s,u,e)$ is the angular frequency of the $j$th atom on the $\ket{p}-\ket{q}$ transition.
In this formula, all terms related to optical frequencies cancel out, which suggests that the optical dephasing has been suppressed by the rephasing pair.
Moreover, by employing the configuration in Fig. \ref{1c} to satisfy the phase matching condition $\bm{k}_{AS}+\bm{k}_{S}=\bm{k}_{w}-\bm{k}_{1}+\bm{k}_{2}+\bm{k}_{r}$ (cf. Appendix \ref{apc}), the spatial phase $\phi_{j0}$ will also vanish in Eq. (\ref{phase}). 
Consequently, the coherence phase is merely determined by the frequencies of spin transitions.
For REICs, the spin inhomogeneous broadening is typically on the order of kHz and such spin dephasing can be further reduced by introducing dynamical decoupling (DD) \cite{jobez2015coherent,laplane2017multimode}. 
Therefore, the effective coherence is rephased. An anti-Stokes photon can be efficiently read out and a strong correlation between the Stokes-anti-Stokes photon pair is expected.

The application of the rephasing pair is the key tool in NLPE-DLCZ.
On the one hand, compared with the AFC-DLCZ scheme, this optical rephasing procedure eliminates the need for complex spectral tailoring and makes full use of the natural absorption.
On the other hand, different from 4L-RASE, the rephasing pair here is composed of a pair of $\pi$ pulses at the same frequency. This enables one to replace these standard $\pi$ pulses with chirped adiabatic control pulses \cite{pascual2013securing,silver1985selective}, which require less laser power and are more robust to the distortion caused by absorption \cite{minavr2010spin, pascual2013securing}.

As we have introduced in Sec. \ref{sec1}, the key to noise suppression of the NLPE protocol is to separate the upper level $\ket{e}$ that emits echoes and the upper level $\ket{u}$ that can be populated by the rephrasing pulse, so that the spontaneous emission noise can be spectrally filtered.
In usual photon-echo based protocols, the imperfection of the $\pi$ pulses will lead to atoms populate $\ket{e}$ and produce indistinguishable spontaneous emission noise.
Differently, in NLPE, most of the residue populations are caused by the imperfection of the rephasing pair, populating in $\ket{u}$. 
The spontaneous emission noise from $\ket{u}$ are spectrally resolvable from the output mode \cite{ma2021elimination}.
Similarly, in the NLPE-DLCZ protocol, the anti-Stokes photons are emitted from the state $\ket{e}$ which is different from the populated excited state $\ket{u}$, ensuring the spontaneous emission noise can be spectrally filtered.
In addition, although the 4L-RASE protocol also separates the upper level $\ket{u}$ that produces RASE and the populated excited state $\ket{e}$, population relaxations from $\ket{e}$ via intermediate crystal field levels during ASE can terminate in $\ket{u}$ (see Fig. \ref{pic0}) and cause indistinguishable spontaneous emission noise \cite{duda2022optimising}.
For NLPE-DLCZ, similar noise issues can be addressed by forbidding the $\ket{s}-\ket{u}$ transition with an applied magnetic field.
A more detailed analysis of the noise will be given in Sec. \ref{sec5}.

The lifetime of the correlation of the photon pairs is limited by the decoherence and spin dephasing.
The decoherence and dephasing of the spin transition $\ket{g}-\ket{s}$ can be reduced by DD during $t\in[T,t_{1}]$ or $t\in[t_{2},t_{r}]$ (cf. Figs. \ref{1a} and \ref{1b}).

Photons resonant with the $\ket{s}-\ket{u}$ transition with a wave vector $\bm{k}'$ could emit at time $t=t_{1}+t_{S}$ due to a four-level echo type rephasing \cite{beavan2011photon}.
However, this emission suffers from the noise issue caused by the incoherent spontaneous emission from $\ket{u}$.
Using the configuration in Fig. \ref{1c}, this invalid emission is silenced due to the phase mismatch $|\bm{k}'|\ll |-\bm{k}_{w}+\bm{k}_{S}+\bm{k}_{1}|$.

\section{Stokes emission}
\label{sec3}
A quantitative analysis of NLPE-DLCZ will be given below. 
We shall model $N$ four-level atoms in a one-dimensional crystal placed in the space $z\in[0,l]$, where $l$ is the length of the crystal and $z$ is the spatial coordinate.
The inhomogeneous broadening of an arbitrary transition $\ket{p}-\ket{q}$ $(p,q =s,g,u,e)$ is characterized by its full width at half maximum (FWHM) $\Gamma_{pq}/(2\pi)$.
The broadenings of transitions $\ket{s}-\ket{e}$ and $\ket{g}-\ket{e}$ are assumed to be the same, $\Gamma_{se}=\Gamma_{ge}=\Gamma$.
The duration of a single temporal mode is defined as $\tau=2\pi/\Gamma$.
The quantum properties of the $j$th atom are described by a set of atomic operators, including the coherence operators $\hat{\sigma}_{pq,j}\equiv\ket{p_{j}}\bra{q_{j}}$ $(p,q=s,g,u,e)$ and the population operators $\hat{\sigma}_{p,j}\equiv\ket{p_{j}}\bra{p_{j}}$ $(p=s,g,u,e)$.

The durations of the short control pulses are ignored in most of the calculations and we use $t^{-}$ ($t^{+}$) to denote the moment immediately before (after) an incident at time $t$.
Photons are described by a light field operator $\hat{\varepsilon}(t,z)$ 
(cf. Appendix \ref{apa}).
It can be divided into two spectral modes $\hat{\varepsilon}_{se}$ and $\hat{\varepsilon}_{ge}$, which are resonant with the transitions $\ket{s}-\ket{e}$ and $\ket{g}-\ket{e}$, respectively.
For the spatial configuration in Fig. \ref{1c}, the Stokes field and the anti-Stokes field are given by $\hat{\varepsilon}_{S}(t)\equiv\hat{\varepsilon}_{se}(t,0)$ and $\hat{\varepsilon}_{AS}(t)\equiv\hat{\varepsilon}_{ge}(t,l)$, respectively. 
Subscripts $S$ and $AS$ always represent Stokes and anti-Stokes in the following.

The interaction between the photon and the optical transition $\ket{p}-\ket{q}$ ($pq=ge,se$) of a single atom can be characterized by a coupling coefficient $\kappa_{pq}$, which is related to the corresponding optical depth $d_{pq}=\alpha_{pq}l$ by $\kappa_{pq}^{2}=d_{pq}/(N\tau)$ (cf. Appendix \ref{apb}). Here $\alpha_{pq}$ is the absorption coefficient.
The reabsorption or gain features of the medium can be characterized by an equivalent absorption function 
\begin{align}
a_{pq}(t,z)=d_{pq}\cdot \frac{1}{N}\sum_{z_{j}<z} [\langle\hat{\sigma}_{p,j}(t)\rangle-\langle\hat{\sigma}_{q,j}(t)\rangle ]
\end{align}

At time $t=0$, the write pulse of duration $\tau$ is sent through the medium, driving the system into \cite{AFC-DLCZ}
\begin{align}
\ket{\psi(0^{+})}=\bigotimes_{j=1}^{N}\ket{\phi_{j}}\otimes\ket{0},
\end{align}
where $\ket{0}$ is the optical vacuum state. The state $\ket{\phi_{j}}$ of the $j$th atom at position $\bm{z}_{j}$ is
\begin{align}
&\ket{\phi_{j}}=G_{j}\ket{g_{j}}+E_{j}\ket{e_{j}},\notag\\
&G_{j}\equiv\cos(\theta_{j}/2), \quad
E_{j}\equiv ie^{i\bm{k}_{w}\cdot\bm{z}_{j}}\sin(\theta_{j}/2),
\end{align}
where $G_{j}$ and $E_{j}$ are coefficients and $\theta_{j}$ is the area of the write pulse felt by the $j$th atom. Considering the attenuation in the crystal, $\theta_{j}=\theta_{0}e^{-\alpha_{ge}z_{j}/2}$.
 
The average probability per unit time of a Stokes emission at time $t$ is given by (cf. Appendix \ref{apa})
\begin{align}
p_{S}(t)=\langle\psi(0^{+})| \hat{\varepsilon}_{S}^{\dag}(t)\hat{\varepsilon}_{S}(t)|\psi(0^{+})\rangle.
\end{align}
Using the general solution (\ref{gen2}) given in Appendix \ref{apb}, the expression of the Stokes field is given by
\begin{align}
\hat{\varepsilon}_{S}(t)=
&\hat{\varepsilon}_{se}(t-l/c,l)e^{a_{se}(0^{+},l)/2}\notag\\
&+i\kappa_{se}\sum_{j=1}^{N}\xi_{j}^{s}(t)\hat{\sigma}_{se,j}(0^{+}),\notag\\
\xi_{j}^{s}(t)\equiv&e^{-i\omega_{se}^{j}t}e^{ik_{se}^{j}z_{j}}e^{a_{se}(0^{+},z_{j})/2},
\label{Sfield}
\end{align}
where $\xi_{j}^{s}$ is a coefficient function for the Stokes field.
The first term of $\hat{\varepsilon}_{S}(t)$ represents the input photons, which stay in the vacuum state.
The second term represents a Stokes emission, where the coherence operator $\hat{\sigma}_{se,j}$ heralds a spin-wave excitation in $\ket{s}$.

The write pulse is weak so that $\theta_{0}$ approaches $0$. To the leading order of $\theta_{0}$, $p_{S}$ is given by (cf. Appendix \ref{apc})
\begin{align}
p_{S}(t_{S})=
\frac{1}{\tau}\frac{\theta_{0}^{2}}{4}\cdot\frac{d_{se}}{d_{ge}}(1-e^{-d_{ge}})=N_{e}d_{se},
\label{Spro}
\end{align}
where $N_{e}=\sum_{j}|E_{j}|^{2}\approx N\theta_{0}^{2}(1-e^{-d_{ge}})/(4d_{ge})$ is the number of atoms excited by the write pulse. 
Although Eq. (\ref{Spro}) is derived for the backward Stokes detection scheme as shown in Fig. \ref{1c}, the formula for that of the forward detection remains the same.

The probability of detecting a double Stokes photon event per mode is approximately $(p_{S}\tau)^{2}$. To avoid such multiple photon events, $p_{S}\ll \tau^{-1}$ should be satisfied in each DLCZ-based protocol \cite{DLCZ}, which can be realized by reducing the intensity of the write pulse or the optical depth $d_{se}$ in NLPE-DLCZ.

\section{anti-Stokes emission}
\label{sec4}
The detailed step-by-step calculation of the readout process introduced in Sec. \ref{sec2} is derived in Appendix \ref{apd}. Here we present the main results.
To the leading order of the coupling constants $\kappa_{ge}$ and $\kappa_{se}$, the anti-Stokes field is given by
\begin{align}
\hat{\varepsilon}_{AS}(t)=
&\hat{\varepsilon}_{ge}(t-l/c,0)e^{a_{ge}(t^{+}_{r},l)/2}\notag\\
&+i\kappa_{ge}\sum_{j}\xi_{j}^{as}(t)\hat{\sigma}_{gs,j}(0^{+}),\notag\\
\xi_{j}^{as}(t)\equiv&ie^{i\delta_{se}^{j}t_{r}-i\omega_{ge}^{j}t+i\delta_{gu}^{j}T}
e^{ik_{ge}^{j}l+i(-\bm{k}_{ge}^{j}-\bm{k}_{1}+\bm{k}_{2}+\bm{k}_{r})\cdot\bm{z}_{j}}\notag\\
&\cdot e^{[a_{ge}(t_{r}^{+},l)-a_{ge}(t_{r}^{+},z_{j})]/2},
\label{Asfield}
\end{align}
where $\delta_{pq}^{j}$ is the angular frequency detuning between the $\ket{p}-\ket{q}$ $(p,q=g,s,u,e)$ transition of the $j$th atom and the control pulse.
$\xi_{j}^{as}$ is a coefficient function for the anti-Stokes field.
The first term of $\hat{\varepsilon}_{AS}(t)$ represents the input field that is in the vacuum state. 
The second term represents an anti-Stokes emission, where the operator $\hat{\sigma}_{gs,j}$ indicates that the anti-Stokes emission is conditioned on the annihilation of a spin-wave excitation.

The average probability per unit time squared to detect a coincidence between a Stokes emission at time $t_{S}$ and an anti-Stokes emission at time $t$ is given by 
\begin{align}
p_{S,AS}(t_{S},t)=
\bra{\psi(0^{+})}\hat{\varepsilon}_{S}^{\dag}(t_{S})\hat{\varepsilon}^{\dag}_{AS}(t)\hat{\varepsilon}_{AS}(t)\hat{\varepsilon}_{S}(t_{S})\ket{\psi(0^{+})}.
\end{align}
If the phase matching condition $\bm{k}_{AS}+\bm{k}_{S}=\bm{k}_{w}-\bm{k}_{1}+\bm{k}_{2}+\bm{k}_{r}$ is satisfied, one obtains
\begin{align}
p_{S,AS}(t_{S},t)=&\frac{1}{\tau}\cdot
p_{S}(t_{S})\cdot \frac{d^{2}_{ge}e^{-d_{ge}}}{1-e^{-d_{ge}}}\notag\\
&\cdot\text{sinc}^2[\pi(t-t_{AS})/\tau]\cdot \mathcal{L}.
\label{coin}
\end{align}
It suggests that the anti-Stokes emission is centered at time $t_{AS}=t_{r}+T-t_{S}$ with a duration $\tau$.
The temporal decays during the rephasing is characterized by a loss parameter
\begin{align}
\mathcal{L}\equiv&\exp[-\tilde{\Gamma}_{gs}^{2}(t_{r}-t_{S})^{2}
-\tilde{\Gamma}_{ue}^{2}T^{2}-2\gamma T\notag\\
&-\gamma_{gs}(t_{AS}-2T)],
\label{L}
\end{align}
which depends on the decoherence and spin dephasing.
The Gaussian spin inhomogeneous broadenings are characterized by the parameters $\tilde{\Gamma}_{gs}=\pi \Gamma_{gs}/\sqrt{2\ln 2}$ and $\tilde{\Gamma}_{ue}=\pi \Gamma_{ue}/\sqrt{2\ln 2}$.
All the optical decoherence rates are assumed to be $\gamma/2$. 
The spin decoherence rate of the $\ket{g}-\ket{s}$ transition is $\gamma_{gs}$.
The spin dephasing of the $\ket{g}-\ket{s}$ transition can be suppressed via an ideal DD, and the temporal decays become 
\begin{align}
\mathcal{L}_{\text{DD}}=\exp[-\tilde{\Gamma}^{2}T^{2}-2\gamma T
-\tilde{\gamma}_{gs}(t_{AS}-2T)].
\label{LDD}
\end{align}
Here $\tilde{\gamma}_{gs}$ is the spin decoherence rate suppressed by DD, and the residual spin dephasing is characterized by $\tilde{\Gamma}^2\equiv\tilde{\Gamma}^{2}_{gs}+\tilde{\Gamma}^{2}_{ue}$.

The readout efficiency $\eta_{\text{N-D}}$ is modeled by the probability of an anti-Stokes emission during $t\in[t_{AS}-\tau/2,t_{AS}+\tau/2]$ conditioned on a Stokes emission during $t\in[t_{S}-\tau/2,t_{S}+\tau/2]$, which is given by
\begin{align}
\eta_{\text{N-D}}=\frac{\tau^2 p_{S,AS}(t_{S},t_{AS})}{\tau p_{S}(t_{S})}
=\frac{d^{2}_{ge}e^{-d_{ge}}}{1-e^{-d_{ge}}}\cdot
\mathcal{L}.
\label{efficiency}
\end{align}
The efficiency formula is independent of $d_{se}$. 
Provided that the temporal decays are ignorable, the upper limit of the efficiency is $65\%$ achieved at $d_{ge}=1.6$.

Eq. (\ref{efficiency}) is derived for the forward anti-Stokes detection scheme as shown in Fig. \ref{1c}. 
Detection of the forward anti-Stokes emission seems to be more feasible in practice and has been adopted by most of the AFC-DLCZ-based experiments \cite{laplane2017multimode,kutluer2017solid,kutluer2019time}.
Despite experimental difficulties, the efficiency formula for backward anti-Stokes detection can be derived following a similar procedure
\begin{align}
\eta_{\text{N-D}}=(1-e^{-d_{ge}}) \cdot\mathcal{L}.
\label{befficiency}
\end{align}
If the temporal decays are ignorable, the efficiency can approach unity as the optical depth grows. This can be understood as the anti-Stokes emission in the backward direction is absorbed less.

\section{noise}
\label{sec5}
Similar to other photon-echo based schemes, major noises in NLPE-DLCZ are caused by the strong control $\pi$ pulses. 
Below, we will expound on their formation mechanisms and offer feasible methods to remove them.

In practice, the absorption of the medium will distort the $\pi$ pulses
and result in incomplete inversions. The relaxation from the residual population on $\ket{u}$ during the anti-Stokes detection will produce direct spontaneous emission noise.
In addition, the strong control pulses will induce free induction decay noise. 
Both of these noise sources can be filtered in the spectrum due to their spectral resolvability \cite{beavan2011photon,ma2021elimination}.

After the inversion of the first $\pi$ pulse, atoms will populate the excited state $\ket{u}$. Then relaxations from $\ket{u}$ may terminate in $\ket{s}$ during $t\in[t^{+}_{1},t^{-}_{2}]$.
In an actual REIC, there exist two relaxation paths: relaxation via the direct optical transition and relaxation through intermediate crystal field states. 
Relaxations through both paths will cause noise issues.
On the one hand, relaxations via the direct optical transition may produce ASE, which will lead to coherent 4L-RASE noises via the rephasing of the second and third $\pi$ pulses.
On the other hand, the populations in $\ket{s}$ from the intermediate state path will be transferred to $\ket{e}$ by the read pulse and produce incoherent spontaneous emission noise.
Both noise sources have the same frequency as the anti-Stokes photons and thus cannot be filtered spectrally. 
The RASE noise can still be eliminated by the configuration in Fig. \ref{1c} via the phase mismatch $|\bm{k}_{2}+\bm{k}_{r}-\bm{k}_{AS}|\gg |\bm{k}_{ASE}|$ \cite{beavan2012demonstration}, where $\bm{k}_{ASE}$ is the wave vector of ASE.
However, eliminating the incoherent noise requires the $\ket{s}-\ket{u}$ transition to be forbidden.

A number of REICs including Eu$^{3+}$:Y$_{2}$SiO$_{5}$ contain non-Kramers ions in noncentrosymmetric crystalline sites, whose crystal field levels are electronic singlets \cite{REICs,engineer}.
For these REICs, a feasible method to forbid the optical transition $\ket{s}-\ket{u}$ is to apply a specific magnetic field \cite{engineer}.
Their $4f^{N}$ wave functions can be separated into electronic and nuclear components \cite{engineer, teplov1967magnetic, PhysRevB.74.195101, MA201832, cruzeiro2018characterization}:
\begin{align}
\ket{g}=\ket{1}\ket{\psi_{g}},\quad \ket{s}=\ket{1}\ket{\psi_{s}},\notag\\
\ket{u}=\ket{2}\ket{\psi_{u}},\quad \ket{e}=\ket{2}\ket{\psi_{e}},
\end{align}
where $\ket{i} (i=1,2)$ and $\psi_{p} (p=g,s,u,e)$ represent the electronic and nuclear hyperfine components, respectively.
Any operator $\hat{o}$ governing the optical transition satisfies
$\langle p|\hat{o}| q\rangle \propto \langle\psi_{p}|\psi_{q}\rangle$, $(p=g,s,q=u,e)$.
Therefore, by tuning the hyperfine interaction with an magnetic field to satisfy the conditions
\begin{align}
\langle\psi_{s}|\psi_{u}\rangle= 0, 
\quad
\langle\psi_{g}|\psi_{e}\rangle, 
\langle\psi_{g}|\psi_{u}\rangle,
\langle\psi_{s}|\psi_{e}\rangle
\neq 0,
\label{conB}
\end{align}
the $\ket{s}-\ket{u}$ transitions can be forbidden \cite{engineer}, while other necessary optical transitions are accessible.
In this way, all noises caused by the relaxations of the $\ket{s}-\ket{u}$ transitions can be eliminated.
Conditions (\ref{conB}) for forbidding a single transition are much looser than those for engineering complete closed optical transitions in Ref. \cite{engineer}.
As a result, there are zero-first-order-Zeeman (ZEFOZ) points \cite{zhong2015optically} that are compatible with conditions (\ref{conB}), which is crucial for creating a long-term correlation.

For example, Refs. \cite{zhong2015optically, ma2021one} have identified a high performance ZEFOZ point of Eu$^{3+}$:Y$_{2}$SiO$_{5}$. 
Following the notations introduced in Ref. \cite{ma2021one}, at a magnetic field magnitude of approximately 1.3 T in the appropriate direction, a dramatic reduction in the decoherence of the hyperfine transition $\ket{3}_{g}-\ket{4}_{g}$ of the electronic ground state $^{7}$F$_{0}$ was realized.
Here, let us consider a NLPE-DLCZ protocol that employs the states $\ket{3}_{g}$ and $\ket{4}_{g}$ as the states $\ket{g}$ and $\ket{s}$, respectively. 
Then we can select the hyperfine states $\ket{1}_{e}$ and $\ket{2}_{e}$ of the electronic state $^{5}$D$_{0}$ as the respective states $\ket{u}$ and $\ket{e}$. Numerical calculation based on the spin Hamiltonian provided in Ref. \cite{cruzeiro2018characterization} gives
\begin{align}
	&|\langle\psi_{s}|\psi_{u}\rangle|=0.01, \quad
	|\langle\psi_{g}|\psi_{e}\rangle|=0.44,\notag\\
	&|\langle\psi_{g}|\psi_{u}\rangle|=0.21, \quad
	|\langle\psi_{s}|\psi_{e}\rangle|=0.29.
\end{align}
Obviously, 
$|\langle\psi_{g}|\psi_{e}\rangle|, |\langle\psi_{g}|\psi_{u}\rangle|, |\langle\psi_{s}|\psi_{e}\rangle|\gg|\langle\psi_{s}|\psi_{u}\rangle|\sim0 $
is obtained at this ZEFOZ field with the proper selection of the upper energy levels, ensuring that the $\ket{s}-\ket{u}$ transition is nearly forbidden while other transitions remains optically accessible.

In addition to the noise caused by the strong rephasing pulse, there is also noise caused by the weak write pulse, which is inherent to any DLCZ-based protocol.
Besides the expected single Stokes emission, the atoms excited by the write pulse will spontaneously decay and emit photons in various temporal-spatial modes.
After interaction with the control pulses, these excitations may terminate in $\ket{e}$ and produce incoherent spontaneous emission. 
The detection of these emissions in the anti-Stokes mode becomes an intrinsic noise source.

Following Ref. \cite{AFC-DLCZ}, we will give an estimate on the upper limit of the intrinsic noise.
As an estimate of the worst case, the decoherence is assumed to be extremely rapid during $t\in[0^{+},t_{r}^{-}]$ and all the atoms excited by the write pulse terminate in $\ket{e}$. 
The atomic system at time $t_{r}^{+}$ is in a complete mixed state, described by the density matrix
\begin{align}
\hat{\rho}(t_{r}^{+})=\bigotimes_{j=1}^{N}
\bigg[
\sum_{p}\langle\hat{\sigma}_{p,j}(t_{r}^{+})\rangle
\ket{p_{j}}\bra{p_{j}}
\bigg]
\otimes\ket{0}\bra{0}.
\end{align}
The rapid decoherence can break any nonclassical correlation of the photon pair and the anti-Stokes field now becomes a noise field
\begin{align}
\hat{\varepsilon}_{n}(t)=&\hat{\varepsilon}_{ge}(t-l/c,0)e^{a_{ge}(t_{r}^{+},l)/2}\notag\\
&+i\kappa_{ge}\sum_{j}\xi_{j}^{n}(t)\hat{\sigma}_{ge,j}(t_{r}^{+}),\notag\\
\xi^{n}_{j}(t)\equiv&e^{-i\omega_{ge}^{j}(t-t_{r})}e^{ik_{ge}^{j}(l-z_{j})}
e^{[a_{ge}(t_{r}^{+},l)-a_{ge}(t_{r}^{+},z_{j})]/2}.
\end{align}
Here $\xi_{j}^{n}$ is a coefficient function for the noise field.
The average probability per unit time to detect a noise photon at time $t$ is equal to
\begin{align}
p_{n}(t)=\text{tr}[\hat{\varepsilon}_{AS}^{\dag}(t)\hat{\varepsilon}_{AS}(t)
\hat{\rho}(t_{r}^{+})]
=\kappa_{ge}^{2}\sum_{j}|\xi_{j}^{n}(t)E_{j}|^{2},
\label{pn}
\end{align}
Then the maximum estimation of the intrinsic noise is given by
\begin{align}
p_{n}(t)\approx \frac{1}{\tau}\cdot \frac{\theta^{2}_{0}}{4}\cdot
d_{ge}e^{-d_{ge}}=\frac{1}{\tau}\frac{N_{e}}{N}\eta^{*}=\frac{p_{S}\cdot \eta^{*}}{d_{se}}.
\end{align}
where $\eta^{*}\equiv d_{ge}^{2}e^{-d_{ge}}/(1-e^{-d_{ge}})$ is the readout efficiency without temporal decays.
Since $\eta^{*}<1$, the number of noise photons per mode is bounded by $p_{n}\tau< N_{e}/N$, where $N_{e}/N$ is the ratio of the atoms excited by the write pulse. As a result, the intrinsic noise can be reduced by a write pulse with a smaller pulse area.

The cross correlation of the Stokes-anti-Stokes photon pair is given by
\begin{align}
g^{(2)}_{S,AS}(t_{S},t)&\approx \frac{p_{S,AS}(t_{S},t)}{p_{S}(t_{S})p_{n}(t)}\notag\\
&=\frac{\eta_{\text{N-D}}}{p_{n}\tau}\cdot\text{sinc}^2[\pi(t-t_{AS})/\tau].
\end{align}
The correlation peak is centered at time $t_{AS}$ with a width $\tau$.
The peak value $g^{2}_{0}\equiv g^{(2)}_{S,AS}(t_{S},t_{AS})$ is actually the signal-to-noise ratio $\eta_{\text{N-D}}/(p_{n}\tau)$, which equals
\begin{align}
g_{0}^{(2)}\approx \frac{4}{\theta_{0}^{2}}\cdot \frac{d_{ge}}{1-e^{-d_{ge}}}\cdot \mathcal{L}
=\frac{d_{se}}{p_{S} \tau}\cdot \mathcal{L},
\label{correlation}
\end{align}
The photon pair can be highly correlated provided that the number of Stokes photons per mode $p_{S}\tau$ is low.
Suppressing the temporal decay $\mathcal{L}$ and increasing the optical depth $d_{se}$ are also beneficial to enhance the correlation.

\section{lifetime and temporal multimode capacity}
\label{sec6}
In this section, we consider the performances of NLPE-DLCZ in the lifetime and the temporal multimode capacity.
The lifetime of the correlation is characterized by the average delay time of the photon pair
\begin{align}
\frac{1}{T}\int_{0}^{T}  (t_{AS}-t_{S}) dt_{S}=t_{r},
\end{align}
The number of temporal modes is roughly $T/\tau=\Gamma T/(2\pi)$.  
To avoid the wrong excitation, $\Gamma$ cannot exceed the angular frequency of the hyperfine transitions, $\Gamma<\omega_{gs},\omega_{ue}$.  
In practice, $\Gamma$ is further restricted to the specific spectral hole burning for initializing atoms in $\ket{g}$ \cite{ma2021elimination}.
Therefore, extending $T$ is the key to improving the temporal multimode capacity.
However, the extension of $T$ is subject to the temporal decays in NLPE-DLCZ.
According to Eqs. (\ref{L}) and (\ref{LDD}), the decoherence and spin dephasing both put limits on $T$.

\begin{figure}[t]
	\centering
	\subfigure[]{
		\includegraphics[width=0.9\linewidth]{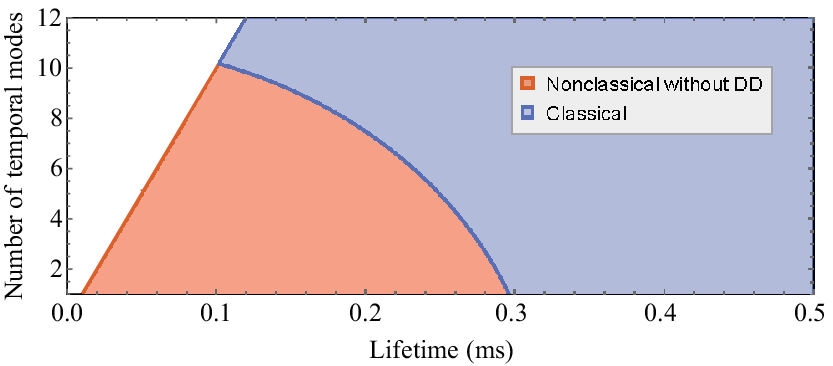}
		\label{2a}
	}
	\subfigure[]{
		\includegraphics[width=0.93\linewidth]{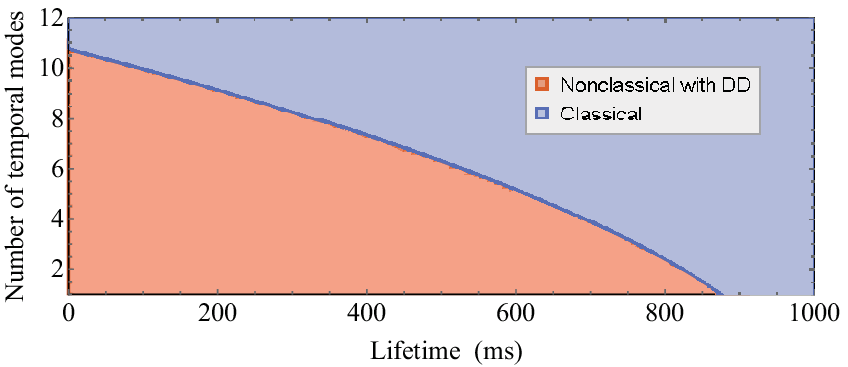}
		\label{2b}
	}
	\caption[]{Feasible lifetime $t_{r}$ and number of temporal modes $T/\tau$ to generate nonclassical correlated photon pairs (a) without dynamical decoupling (DD) or (b) with DD.
	The red regions represent the feasible values of $t_{r}$ and $T/\tau$, where all photon pairs are nonclassically correlated.
	The DD process significantly extends the lifetime of the correlation and helps to enlarge the temporal capacity.
	In the blue regions, some of the photon pairs are classically correlated.
	Both curved boundaries of the noncalssical regions represent $g^{(2)}_{0}=2$, which are conic sections.
	The straight boundaries represent $T=0$ and $t_{r}=2T$.
	These parameters are $\gamma_{gs}=50$ Hz, $\tilde{\gamma}_{gs}=2.5$ Hz, $\tilde{\Gamma}_{gs}=5$ kHz, $\tilde{\Gamma}_{ue}=20$ kHz, $\gamma=10$ kHz, $\tau=5$ $\mu$s, $d_{se}=1$ and $p_{S}=0.01$ $\mu$s$^{-1}$. 
	The plot parameters are typical values for Eu$^{3+}$:Y$_{2}$SiO$_{5}$ at zero field \cite{laplane2017multimode,ma2021elimination,arcangeli2014spectroscopy}.
	}
	\label{pic2}
\end{figure}

The criterion of the noncalssical feature of the correlation is $g_{0}^{(2)}>2$,
which can be deduced from the violation of the Cauchy-Schwartz inequality \cite{chou2004single}.
This criterion also gives a quantitative estimate of both the lifetime $t_{r}$ and the temporal mode capacity $T/\tau$. 
Let us consider the case without DD first. Using Eq. (\ref{correlation}), the criterion is written as
\begin{align}
\mathcal{L}>2p_{S}\tau/d_{se}.
\end{align}
To ensure that all the photon pairs are nonclassically correlated, this inequality should hold for Stokes photons emitted at any possible instant $t_{S}\in[0,T]$.
Thus, the complete restrictions of $t_{r}$ and $T$ are given by:
\begin{align}
&\tilde{\Gamma}_{gs}^2 t_{r}^2+\gamma_{gs}t_{r}
+\tilde{\Gamma}_{ue}^2 T^{2}+(2\gamma-\gamma_{gs}) T<\ln(d_{se}/2p_{S}\tau),
\notag\\
&T>0, \quad t_{r}>2T.
\label{res}
\end{align}
Here, the first inequality arises the violation of the Cauchy-Schwarz inequality, while the other inequalities are the natural requests of the pulse sequence. 
The solution region is shown in Fig. \ref{2a}.
The maximum values of $t_{r}$ and $T/\tau$ are
\begin{align}
&(t_{r})_{\text{Max}}\approx 
\frac{\sqrt{\ln[d_{se}/(2p_{S}\tau)]}}{\tilde{\Gamma}_{gs}},\notag\\
&(T/\tau)_{\text{Max}}\approx
\frac{-\gamma
	+\sqrt{\gamma^2+(\tilde{\Gamma}_{ue}^2+4\tilde{\Gamma}_{gs}^2)
		\ln[d_{se}/(2p_{S}\tau)]}}{(\tilde{\Gamma}_{ue}^2+4\tilde{\Gamma}_{gs}^2)\tau}.
\end{align}
These two maxima cannot be obtained simultaneously since there is a trade-off between the lifetime and the temporal multimode capacity, as shown by Fig. \ref{2a}.

The results with DD can be similarly derived by replacing $\mathcal{L}$ with $\mathcal{L}_{\text{DD}}$. The restrictions set is written
\begin{align}
&\tilde{\gamma}_{gs}t_{r}
+\tilde{\Gamma}^2 T^{2}+(2\gamma-\tilde{\gamma}_{gs}) T<\ln(d_{se}/2p_{S}\tau),
\label{resdd}\notag\\
&T>0, \quad t_{r}>2T.
\end{align}
The solution region is also depicted in Fig. \ref{pic2}.
The maximum values of $t_{r}$ and $T/\tau$ are then
\begin{align}
&(t_{r})_{\text{Max}}\approx  \frac{\ln(d_{se}/2p_{S}\tau)}{\tilde{\gamma}_{gs}},\notag\\
&(T/\tau)_{\text{Max}}\approx
\frac{-\gamma +\sqrt{\gamma^2+\tilde{\Gamma}^{2}\ln(d_{se}/2p_{S}\tau) }}{\tilde{\Gamma}^2\tau}.
\end{align}
Similarly, these maximum values cannot be achieved in the same trial.

The solution regions in Fig. \ref{2b} show that a powerful DD greatly extends the lifetime of the correlations and helps to enlarge the temporal multi-mode capacity.
Yet, due to various temporal decays, the NLPE-DLCZ scheme has a less temporal multimode capacity compared to that in the AFC-DLCZ scheme. Alternatively, one could implement multiplexing in other degrees of freedom, such as spectral \cite{sinclair2014spectral,seri2019quantum} and spatial multiplexing \cite{yang2018multiplexed,zhou2015quantum}, to enhance the multimode capacity.

\section{applications}
\label{sec7}
\subsection{Comparison with Existing Protocols}
To determine the practical applications of NLPE-DLCZ, we will make a comprehensive comparison between NLPE-DLCZ and existing schemes in this section.
Let us start with the comparison with AFC-DLCZ.
Ref. \cite{AFC-DLCZ} has given the efficiency formula of AFC-DLCZ
\begin{align}
&\eta_{\text{A-D}}=\frac{\tilde{d}_{ge}^2 e^{-\tilde{d}_{ge}}}{1-e^{-\tilde{d}_{ge}}} \cdot e^{-\frac{2\pi\mathcal{K}^2}{F^2}}\cdot \mathcal{L}'\notag\\
&\mathcal{L}'\equiv\exp[-\tilde{\Gamma}_{gs}^{2}(t_{r}-t)^{2}-\gamma T-\gamma_{gs} (t_{AS}-T)]
\label{afcef}
\end{align}
Here the temporal decay term $\mathcal{L}'$ is added to the original formula in Ref. \cite{AFC-DLCZ}.
$\tilde{d}_{ge}=\mathcal{K} d_{ge}/F$ is the effective optical depth and $F$ is the finesse. The constant $\mathcal{K}\equiv\sqrt{\pi/(4\ln2)}$ characterizes the frequency comb made with Gaussian peaks.

\begin{figure}[t]
	\centering
	\subfigure[]{
		\includegraphics[width=0.95\linewidth]{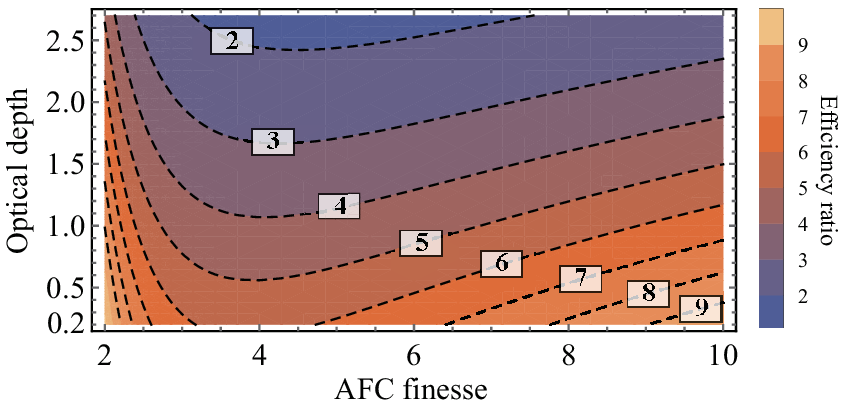}
		\label{3a}
	}
	\subfigure[]{
		\includegraphics[width=0.95\linewidth]{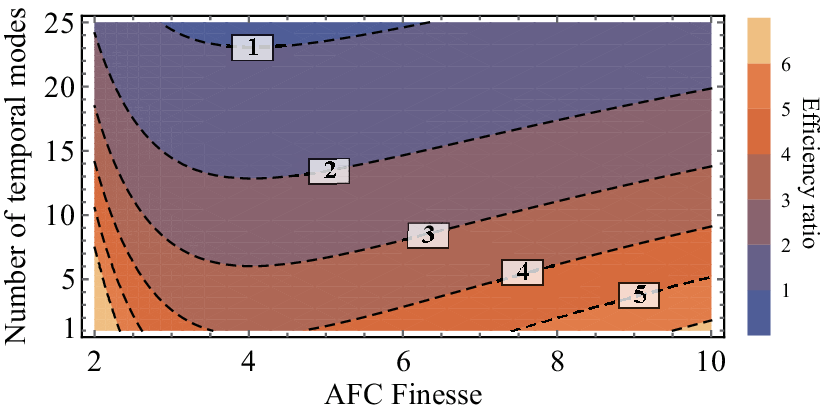}
		\label{3b}
	}
	\caption{The efficiency ratio $\mathcal{R}$ of NLPE-DLCZ and AFC-DLCZ as a function of (a) the AFC finesse $F$ and the optical depth $d_{ge}$ or (b) the AFC finesse and the number of temporal modes.
	In (a) we consider the case of low temporal capacity and assume that the temporal decays of the two schemes are close, $\mathcal{L}/\mathcal{L'}\approx 1$.
	The contour map in (a) shows a significant efficiency improvement of NLPE-DLCZ in the regime $d_{se}<2.5$, ranging from 2 times to more than 9 times.
	In (b) we fix the optical depth to $d_{ge}=1$ and consider temporal multimode capacity. The contour map in (b) indicates that below 25 temporal modes NLPE-DLCZ can achieve an efficiency enhancement.
	The plot parameters of (b) are the same as those in Fig. \ref{pic2}.
	}
	\label{pic3}
\end{figure}

We first concentrate on the low optical depth regime, $d_{ge},d_{se}<1.5$, 
which is typical for Eu$^{3+}$:Y$_{2}$SiO$_{5}$ in a strong magnetic field.
The efficiency formulas (\ref{efficiency}) and (\ref{afcef}) can be expanded in terms of $d_{ge}$.
Omitting high-order terms, one obtains an rough estimate of the respective efficiencies of the AFC-DLCZ and NLPE-DLCZ schemes
\begin{align}
\eta_{\text{A-D}}\approx\tilde{d}_{ge}e^{-\frac{2\pi\mathcal{K}^{2}}{F^2}}\mathcal{L}', \quad
\eta_{\text{N-D}}\approx d_{ge}\mathcal{L}.
\end{align}
The ratio of the efficiencies $\mathcal{R}=\eta_{\text{N-D}}/\eta_{\text{A-D}}$ equals
\begin{align}
\mathcal{R}
\approx (F/\mathcal{K}) e^{\frac{2\pi\mathcal{K}^2}{F^2}}
\cdot e^{-\tilde{\Gamma}_{ue}^{2}T^{2}-\gamma T}.
\label{ratio}
\end{align}
In applications with low temporal multimode capacity requirements, the term related to $T$ can be neglected and $\mathcal{R}>6$ is obtained, which means that the efficiency of NLPE-DLCZ can be enhanced by 6 times compared to that of AFC-DLCZ.
A precise efficiency comparison is depicted in Fig. \ref{pic3}.
For single temporal mode correlation, Fig. \ref{3a} shows an obvious efficiency enhancement of NLPE-DLCZ in the low optical depth regime, when compared with the high finesse AFC-DLCZ.  
In Fig. \ref{3b}, the temporal multimode correlation is considered and the optical depth is fixed at $d_{ge}=1$.
It suggests that below 25 temporal modes, NLPE-DLCZ can maintain its efficiency advantage.

The efficiency improvement of NLPE-DLCZ is beneficial to the creation of highly correlated photon pairs.
Quantitatively, in the low optical depth regime, the maximum value of the correlation peak for AFC-DLCZ can be derived from Ref. \cite{AFC-DLCZ}
\begin{align}
{g'}_{0}^{(2)}\approx \frac{d_{se}}{p_{S}\tau}\cdot\frac{\mathcal{K}}{F} e^{-\frac{2\pi\mathcal{K}^{2}}{F^2}}\mathcal{L'},
\end{align}
which is related to the peak value of NLPE-DLCZ by $g_{0}^{(2)}/{g'}^{(2)}_{0}=\mathcal{R}$ according to Eqs. (\ref{correlation}) and (\ref{ratio}). In other words, the ratio of the cross-correlation values of the two schemes is the same as the ratio of the efficiencies.
Hence, an enhanced cross-correlation could be expected with NLPE-DLCZ in the low-absorption regime.

The weakness of NLPE-DLCZ compared with AFC-DLCZ is the temporal multimode capacity. As $T$ increases, both the efficiency and correlation enhancements continue to decrease.
In addition, in the optically dense regime, AFC-DLCZ could approach its maximum efficiency and NLPE-DLCZ has no advantage in this regime.

We further compare NLPE-DLCZ with RASE. 
The two-level RASE is sensitive to the distortions of the $\pi$ pulses and faces a trade-off between the readout efficiency and the single photon purity \cite{Stevenson2014}, so only the 4L-RASE is considered here.
Since the temporal decays are similar for both schemes, for simplicity all the temporal decays are neglected in the following discussions.
The readout efficiency of 4L-RASE can be calculated by both numerical \cite{Stevenson2014} and analytical \cite{RASE,duda2022optimising} methods.
In the low optical depth regime and provided ideal $\pi$ pulses, the efficiency of 4L-RASE is approximately $\eta_{\text{RASE}}\approx d_{ge}$, which is the same as that of NLPE-DLCZ.
However, standard $\pi$ pulses will be distorted by absorption in practice, which reduces the efficiency. 
In NLPE-DLCZ, the rephasing pair can be composed of a pair of chirped adiabatic pulses to increase the robustness to absorption and enhance the efficiency.
This strategy cannot be realized in 4L-RASE since the frequencies of the $\pi$ pulses are different there.

Another improvement is the rejection of certain noise.
In 4L-RASE, indirect decays via the intermediate crystal field levels and the off-resonant excitation by the inversion pulse will populate $\ket{s}$ and cause incoherent spontaneous emission noises \cite{beavan2012demonstration,duda2022optimising}.
In NLPE-DLCZ, similar noise sources can be eliminated by forbidding the transition with an applied magnetic field.
However, approaches based on forbidding transition cannot be implemented in 4L-RASE, because the corresponding transition has been used to produce the ASE.

Since the two protocols process similar temporal decays and a slightly higher correlation could be created by NLPE-DLCZ, one can expect that NLPE-DLCZ may achieve a slightly better lifetime and temporal multimode capacity.

The above discussion is for the low optical depth regime.
In the optically dense regime, numerical calculations indicate that 4L-RASE has a higher efficiency ceiling ($\approx 70\%$ \cite{Stevenson2014}) than that of NLPE-DLCZ.

\subsection{Experiment Outlook}
We have shown that NLPE-DLCZ could serve as a correlated photon pair source with good performances in the optically thin regime of REICs. 
In this subsection, we will provide outlooks on its experimental applications.

Implementing a high-quality DLCZ based protocol with Eu$^{3+}$: Y$_{2}$SiO$_{5}$ presents a significant challenge \cite{laplane2017multimode}, mainly due to the trade-off between a long spin coherence time and a large optical depth within a single crystal \cite{ma2023monte}.
NLPE-DLCZ could enable the generation of high quality correlated photon pairs based on Eu$^{3+}$: Y$_{2}$SiO$_{5}$ while preserving the long coherence time of this material.
Thus, the primary goal of the NLPE-DLCZ protocol is to serve as a component of the DLCZ-based quantum communications with Eu$^{3+}$: Y$_{2}$SiO$_{5}$. 
One the one hand, NLPE-DLCZ could be used to construct a temporal multi-mode quantum repeater using Eu$^{3+}$: Y$_{2}$SiO$_{5}$ \cite{PhysRevLett.98.190503}. 
By preserving more natural absorption, it might achieve an efficiency improvement compared to AFC-DLCZ.
On the other hand, when operating at ZEFOZ fields, NLPE-DLCZ could enable the demonstrations of transportable quantum memories for entanglements  \cite{kutluer2019time} with Eu$^{3+}$: Y$_{2}$SiO$_{5}$ \cite{morton2015spin}.
The slow nuclear spin decoherence of Eu$^{3+}$: Y$_{2}$SiO$_{5}$ enables the long storage time of the photon pairs, which directly determines the distance that the memory can be transported.

From a general perspective, the pulse sequence of NLPE-DLCZ or AFC-DLCZ is nearly identical to the original DLCZ, except that an extra physical process or structure is added for rephasing.
This suggests that corresponding to different rephasing techniques there exist many other echo-based DLCZ schemes that can be developed for inhomogeneously broadened systems.

\section{conclusion}
\label{sec8}
In conclusion, we propose a DLCZ-like scheme based on the noiseless photon echo technique. 
The NLPE-DLCZ scheme is developed for inhomogeneously broadened four-level atomic systems, especially for solid-state medium containing rare-earth ions.
The dephasing problem is solved by a pair of identical $\pi$ pulses, which can also be replaced by rubost chirped adiabatic pulses.
Photon pairs with strong cross-correlations can be obtained even with optically thin mediums. Therefore, the NLPE-DLCZ scheme is particularly suitable for implementations in Eu$^{3+}$:Y$_{2}$SiO$_{5}$ which is an important material for quantum networking \cite{liu2021heralded,lago2021telecom,ortu2022storage,PhysRevLett.125.260504,PhysRevLett.128.180501} and transportable quantum memories \cite{zhong2015optically,ma2021one}.

\begin{acknowledgments}
This work is supported by the National Key R\&D Program of China (No. 2017YFA0304100), Innovation Program for Quantum Science and Technology (No. 2021ZD0301200), the National Natural Science Foundation of China (Nos. 12222411, 11774335 and 11821404) and the Fundamental Research Funds for the Central Universities (No. WK2470000026 and No. WK2470000029). Z.-Q.Z acknowledges the support from the Youth Innovation Promotion Association CAS.
\end{acknowledgments}

\appendix

\section{Number of photons}
\label{apa}
This appendix describes the number of photons in terms of the photon field operators.
We define the one-dimensional light field $\hat{\varepsilon}(t,z)$ as
\begin{align}
\hat{\varepsilon}(t,z)=
\frac{1}{\sqrt{2\pi/c}}
\int_{\infty} dk 
\hat{a}_{k}(t)e^{ik z}.
\end{align}
Here $c$ is the speed of light. $\hat{a}_{\omega}$ is the photon annihilation operator and the commutation relations are written as
\begin{align}
&[\hat{a}_{k}(t),\hat{a}^{\dag}_{k'}(t)]=\delta(k-k'),\notag\\
&[\hat{\varepsilon}(t,z),\hat{\varepsilon}^{\dag}(t,z')]=
c\delta(z-z').
\end{align}

The total number of photons at time $t$ is described by an operator $\hat{\mu}(t)$. 
In the frequency space, $\hat{\mu}(t)$ and $\hat{a}_{\omega}$ are related by
\begin{align}
\hat{\mu}(t)=\int_{\infty} \hat{a}^{\dag}_{k}(t)\hat{a}_{k}(t) dk.
\end{align}
In the configuration space, by integrating $\hat{\varepsilon}$, one obtains
\begin{align}
\int_{-\infty}^{\infty} c^{-1}\hat{\varepsilon}^{\dag}\hat{\varepsilon}(t,z) dz
=\int_{\infty} dk dk' \hat{a}_{k}^{\dag}\hat{a}_{k'}
\delta(k-k')
=\hat{\mu}(t).
\label{A4}
\end{align}
This suggests that $c^{-1}\hat{\varepsilon}^{\dag}\hat{\varepsilon}(t,z)$ is the spatial density of photons. The number of photons in a local area $z\in[a,b]$ is given by
\begin{align}
\hat{\mu}_{[a,b]}(t)=
\int_{a}^{b} c^{-1}\hat{\varepsilon}^{\dag}\hat{\varepsilon}(t,z) dz
\end{align}

The number of photons traveling forward through the position $z=l$ until time $t$ is written as $\hat{\mu}_{o}(t)=\hat{\mu}_{[l,\infty]}(t)$.
Provided that the photodetector is placed far away (at $z=+\infty$),
in the space $z\in[l,\infty]$ the photon field is written as
\begin{align}
\hat{\varepsilon}(t,z)=\hat{\varepsilon}[t-(z-l)/c,l].
\end{align}
This yields a new expression of output photons in terms of the time integral:
\begin{align}
\hat{\mu}_{o}(t)=\int_{-\infty}^{t} \hat{\varepsilon}^{\dag}\hat{\varepsilon}(t',l) dt'.
\end{align}
It reveals that the operator $\hat{\varepsilon}^{\dag}\hat{\varepsilon}(t,l)$ is the time derivative of the output photons, which describes the photons traveling forward through the position $z=l$ per unit time.
The total number of these photons detected in a time window $T_{s}$ is given by
\begin{align}
\hat{\mu}_{o}
=\int_{T_{s}} \hat{\varepsilon}^{\dag}\hat{\varepsilon}(t,l) dt.
\end{align}
The expression for photons traveling backward through the position $z=0$ is similar:
\begin{align}
\hat{\mu}_{o}
=\int_{T_{s}} \hat{\varepsilon}^{\dag}\hat{\varepsilon}(t,0) dt.
\end{align}

\section{General equations}
\label{apb}
Here, we will derive and solve the general equation of motions.
A four-level system can be decomposed into several two-level subsystems. It is sufficient to restrict the framework of the analysis to a two-level subsystem. Hence, we shall model an ensemble of two-level atoms here. 
The lower and upper energy levels are denoted by $\downarrow$ and $\uparrow$, respectively. 
The $j$th atom is described by several atomic operators
\begin{align}
\hat{\sigma}_{j}=\ket{\downarrow}_{jj}\bra{\uparrow},
\quad
\hat{\sigma}_{z,j}=\ket{\uparrow}_{jj}\bra{\uparrow}-\ket{\downarrow}_{jj}\bra{\downarrow}.
\end{align}
Its dipole moment can be expanded to $\hat{\bm{d}}_{j}=\bm{d}(\hat{\sigma}_{j}+\hat{\sigma}_{j}^{\dag})$. For simplicity, the transition dipole moment $\bm{d}$ is considered to be a uniform real vector.

The resonant electric field of photons is written as 
\begin{align}
\hat{\bm{E}}(t,z)=\frac{\hbar\kappa \bm{\epsilon}}{(\bm{d}\cdot\bm{\epsilon})}
\hat{\varepsilon}(t,z)+H.c.
\end{align}
Here $\bm{\epsilon}$ is the polarization vector.
$\kappa=\bm{d}\cdot\bm{\epsilon}\sqrt{\omega_{0}/(2\hbar c\epsilon_{0}A)}$ characterizes the coupling strength, where $\omega_{0}$ is the central angular frequency of the light and $A$ is the beam cross sectional area. 

The Hamiltonian of the system is given by 
\begin{align}
\hat{H}=&\int \hbar\omega \hat{a}^{\dag}_{k}\hat{a}_{k}dk+
\sum_{j} \hbar 
\omega_{j}\hat{\sigma}^{\dag}_{j}\hat{\sigma}_{j}\notag\\
&-\hbar\kappa  \sum_{j}
\hat{\sigma}^{\dag}_{j}(t)\hat{\varepsilon}(t,z_{j})+H.c.,
\end{align}
where $\omega_{j}=\omega_{0}+\delta_{j}$ is the angular frequency of the $j$th atom. $\delta_{j}$ is the detuning.
Then the Heisenberg equations yield
\begin{align}
\pm\partial_{z}\hat{\varepsilon}+c^{-1}\partial_{t}\hat{\varepsilon} &
=i\kappa \sum_{j}\delta(z-z_{j})
\hat{\sigma}_{j},\notag\\
\partial_{t}\hat{\sigma}_{j}&=-i\omega_{j}\hat{\sigma}_{j}
-i\kappa\hat{\sigma}_{z,j}\hat{\varepsilon}_{j},
\label{gen0}
\end{align}
where we denote $\hat{\varepsilon}_{j}\equiv \hat{\varepsilon}(t,z_{j})$. 
The value of $\pm$ is $+$ for forward propagation and $-$ for backward propagation.

Due to the nonlinear coupling term $\hat{\sigma}_{z,j}\hat{\varepsilon}_{j}$, Eqs. (\ref{gen0}) are difficult to solve directly.
Note that $\hat{\sigma}_{z,j}$ has two eigenstates $\ket{e}_{j}$ and $\ket{g}_{j}$.
When the atomic system approaches the eigenstate $\prod_{j}\ket{g_{j}}$ or $\prod_{j}\ket{e_{j}}$, the operator $\hat{\sigma}_{z,j}(t)$
can be approximately replaced by its initial average \cite{afzelius2009,AFC-DLCZ,RASE}
\begin{align}
\hat{\sigma}_{z,j}(t)\approx \langle\hat{\sigma}_{z,j}(t)\rangle\approx 
\langle\hat{\sigma}_{z,j}(t_{0})\rangle,
\label{appro}
\end{align}
where $t_{0}$ is the initial instant.
Considering that atoms are evenly distributed over an inhomogeneous broadening with a FWHM equal to $\Gamma/(2\pi)$ and using the approximation (\ref{appro}), Eqs. (\ref{gen0}) is reduced to 
\begin{align}
(\pm\partial_{z}+c^{-1}&\partial_{t})\hat{\varepsilon}(t,z)
=\frac{1}{2}\alpha \langle\hat{\sigma}_{z,j|z_{j}=z}(t_{0})\rangle \hat{\varepsilon}(t,z)\notag\\
&+i\kappa \sum_{j}\delta(z-z_{j})\hat{\sigma}_{j}(t_{0})
e^{-i\omega_{j}(t-t_{0})},
\label{eqn}
\end{align}
where 
$\alpha=2\pi\kappa^2 N/(l\Gamma)$ is the absorption coefficient.

Linear equation (\ref{eqn}) is solvable. 
The solution for the forward field is given by
\begin{align}
\hat{\varepsilon}(t,z)&=\hat{\varepsilon}(t-z/c,0)e^{a(z)/2}\notag\\
+&\sum_{z_{j}<z}
i\kappa\hat{\sigma}_{j}(t_{0})
e^{-i\omega_{j}(t-t_{0})}
e^{ik_{j}(z-z_{j})}
e^{[a(z)-a(z_{j})]/2},
\label{gen1}
\end{align}
while the solution for the backward field is given by
\begin{align}
\hat{\varepsilon}&(t,z)=\hat{\varepsilon}[t-(l-z)/c,l]e^{[a(l)-a(z)]/2}\notag\\
&+ \sum_{z<z_{j}<l}
i\kappa\hat{\sigma}_{j}(t_{0})
e^{-i\omega_{j}(t-t_{0})}
e^{-ik_{j}(z-z_{j})}
e^{-[a(z)-a(z_{j})]/2}.
\label{gen2}
\end{align}
Here $k_{j}\equiv\omega_{j}/c$.
The equivalent absorption function $a(z)$ is defined as
\begin{align}
a(z)\equiv d \cdot \frac{1}{N} \sum_{z_{j}<z} \langle\hat{\sigma}_{z,j}(t_{0})\rangle.
\end{align}

\section{Emission of the Stokes photons}
\label{apc}
Here we derive formulas for the Stokes emission.
According to the output Stokes field Eq. (\ref{Sfield}), the Stokes detection probability is written as
\begin{align}
p_{S}(t)=\kappa^{2}_{se} \sum_{j} |\xi^{s}_{j}(t) E_{j}|^2.
\end{align}
Replacing the summation with the integral, one obtains
\begin{align}
p_{S}(t)
=\frac{1}{\tau} \cdot \alpha_{se} \int_{0}^{l} e^{a_{se}(0^{+},z)} \sin^{2}[\theta(z)/2] dz.
\end{align}
Here $\theta(z)=\theta_{j|z_{j}=z}$ is the optical area of the write pulse at position $z$.
Substituting $\theta(z)=\theta_{0}e^{-\alpha_{ge} z/2}$, $a_{se}(z)$ can be expanded to
\begin{align}
a_{se}(0^{+},z)= 
\frac{\alpha_{se}}{\alpha_{ge}}(1-e^{-\alpha_{ge}z})\theta^2_{0}
+o(\theta_{0}^{4}).\quad (\theta_{0}\approx 0)
\end{align}
Hence, the Stokes detection probability for backward detections is given by
\begin{align}
p_{S}=\frac{1}{\tau}\cdot
\frac{\theta_{0}^{2}}{4}\frac{d_{se}}{d_{ge}}(1-e^{-d_{ge}})
+o(\theta_{0}^{4}).
\end{align}
It is easy to verify that the Stokes detection probability for forward detections is the same.

\section{Coincidence of Stokes-anti-Stokes pairs}
\label{apd}
This appendix continues to calculate the coincidence detection probability $p_{S,AS}$. 
The first step is to calculate $\hat{\varepsilon}_{AS}(t)\hat{\varepsilon}_{S}(t_{S})\ket{\psi(0^{+})}\notag$. The anti-Stokes field can be obtained from the solution (\ref{gen1})
\begin{align}
\hat{\varepsilon}_{AS}(t)=&\hat{\varepsilon}_{ge}(t-l/c,0)e^{a_{ge}(t_{r}^{+},l)/2}\notag\\
+&i\kappa_{ge} \sum_{j} \xi^{n}_{j}(t_{r}^{+})
\hat{\sigma}_{ge,j}(t_{r}^{+}).
\label{as}
\end{align}
The evolution of the coherence $\hat{\sigma}_{ge,j}(t_{r}^{+})$ via the pulse sequence of NLPE-DLCZ can be solved by the optical Bloch equations \cite{scully1999quantum}
\begin{align}
&\hat{\sigma}_{ge,j}(t_{r}^{+})=ie^{i(\phi_{j}+\varphi_{j})}\hat{\sigma}_{gs,j}(0^{+})
+o(\kappa_{pq})\notag\\
&\phi_{j}=\bm{k}_{r}\cdot\bm{z}_{j}-\omega_{gs}^{j}t_{r}-\omega_{se}t_{r},
\notag\\
&\varphi_{j}=(-\bm{k}_{1}+\bm{k}_{2})\cdot\bm{z}_{j}+(\omega_{gu}^{j}-\omega_{gu})T.
\label{coh}
\end{align}
Here, $\omega_{ge}$ is the central frequency of the write pulse.
$\phi_{j}$ is the phase increment involving the write and read pulses, 
while $\varphi_{j}$ is related to the rephasing pair.
To the leading order of $\kappa_{ge}$ and $\kappa_{se}$, by plugging Eq. (\ref{coh}) into Eq. (\ref{as}), we obtain Eq. (\ref{Asfield}).
The omitted term $o(\kappa_{pq})$ ($pq=ge,se$) actually accounts for some spectrally resolvable noise such as the free induced decays, which is beyond our interests.

Eqs. (\ref{Sfield}) and (\ref{Asfield}) reduce the projection to
\begin{align}
	&\hat{\varepsilon}_{AS}(t)\hat{\varepsilon}_{S}(t_{S})\ket{\psi(0^{+})}\notag\\
	&=-\kappa_{ge}\kappa_{se}\sum_{j}\xi_{j}^{as}(t)\xi_{j}^{s}(t_{S}) E_{j}
	\bigg[\ket{g_{j}}\bigotimes_{j'\neq j}\ket{\phi_{j'}}\bigg]\otimes\ket{0}.
\end{align}
Then the expression of the coincidence detection probability is 
\begin{align*}
p_{S,AS}=&(\kappa_{ge}\kappa_{se})^2 \sum_{j}|\xi^{as}_{j}(t)\xi^{s}_{j}(t_{S})E^{2}_{j}|^2 \notag\\
&+(\kappa_{ge}\kappa_{se})^2\bigg|\sum_{j}\xi^{as}_{j}(t)\xi^{s}_{j}(t_{S})
E_{j}G_{j}\bigg|^{2}.
\end{align*}
The first term of $p_{S,AS}$ can be further reduced to
\begin{align*}
	&(\kappa_{ge}\kappa_{se})^2 \sum_{j}|\xi^{as}_{j}(t)\xi^{s}_{j}(t_{S})E^{2}_{j}|^2\notag\\
	=&\frac{1}{\tau^{2}} \frac{\alpha_{ge}\alpha_{se}l}{N}\int_{0}^{l}
	e^{a_{se}(0^{+},z)+a_{ge}(t_{r}^{+},l)-a_{ge}(t_{r}^{+},z)}\sin^{4}(\theta/2) dz.
\end{align*}
It is an extremely small amount proportional to $N^{-1}$ and thus can be omitted, so that 
\begin{align}
p_{S,AS}(t_{S},t)=(\kappa_{ge}\kappa_{se})^2\bigg|\sum_{j}\xi^{as}_{j}(t)\xi^{s}_{j}(t_{S})E_{j}G_{j}\bigg|^{2}.
\end{align}
Replacing the summation with integral one obtains
\begin{align}
p_{S,AS}=\frac{1}{\tau^{2}}\cdot \Pi_{S,AS}(t_{S},t)
\cdot \zeta_{S,AS}(d_{ge},d_{se},\theta_{0}).
\end{align}
The temporal term $\Pi_{S,AS}$ is
\begin{align}
\Pi_{S,AS}(t_{S},t)
=&\bigg|\int_{\Gamma_{ge}} \rho_{ge}(\Delta_{ge}) 
e^{-i\Delta_{ge}(t_{S}+t-t_{r}-T)} d\Delta_{ge}\bigg|^2
\notag\\
&\cdot \bigg|\int_{\gamma_{sg}}\rho_{sg}(\delta_{sg})
  e^{i\delta_{sg}(t_{r}-t_{S})}d\delta_{sg}\bigg|^2\notag\\
&\cdot\bigg|\int_{\gamma_{ue}}\rho_{ue}(\delta_{ue})
e^{-i\delta_{ue}T}d\delta_{ue}\bigg|^2,
\end{align}
where $\rho_{ge}(\Delta_{ge})$, $\rho_{sg}(\delta_{sg})$ and $\rho_{ue}(\delta_{ue})$ is the normalized angular frequency density.
We adopt the same inhomogeneous broadening configuration as in Ref. \cite{ma2021elimination} and obtain
\begin{align}
\Pi_{S,AS}(t_{S},t)=
\text{sinc}^2[\pi(t-t_{AS})/\tau]\mathcal{L}.
\end{align}
The optical depth term $\zeta_{S,AS}$ is
\begin{align}
&\zeta_{S,AS}(d_{ge},d_{se},\theta_{0})
=\alpha_{ge}\alpha_{se}\bigg|\int_{0}^{l}
e^{i(\bm{k}_{w}-\bm{k}_{S}-\bm{k}_{1}+\bm{k}_{2}+\bm{k}_{r}-\bm{k}_{AS})\cdot\bm{z}}\notag\\
&\cdot \frac{1}{2} e^{[a_{se}(0^{+},z)+a_{ge}(t_{r}^{+},l)-a_{ge}(t_{r}^{+},z)]/2} \sin[\theta(z)] dz\bigg|^{2}.
\end{align}
It requires the phase matching condition 
\begin{align}
\bm{k}_{w}-\bm{k}_{S}-\bm{k}_{1}+\bm{k}_{2}+\bm{k}_{r}-\bm{k}_{AS}=0.
\end{align}
to make the integral valid. $a_{ge}(t_{r}^{+},z)$ is given by
\begin{align}
a_{ge}(t_{r}^{+},z)=-\alpha_{ge} z+(1-e^{-\alpha_{ge}z})\theta_{0}^{2}+o(\theta_{0}^{4}).
\end{align}
To the leading order of $\theta_{0}$, one obtains 
\begin{align}
\zeta_{S,AS}(d_{ge},d_{se},\theta_{0})=\frac{\theta_{0}^{2}}{4} d_{ge}d_{se}e^{-d_{ge}}+o(\theta_{0}^{2}).
\end{align}
Eventually, the coincidence detection probability is given by Eq. (\ref{coin}).


%

\end{document}